\documentclass[twocolumn,prl,aps,footinbib,amsmath,amssymb,superscriptaddress,longbibliography]{revtex4-1}
\usepackage{graphicx}
\usepackage{dcolumn}
\usepackage{xcolor}
\usepackage{hyperref}
\usepackage{natbib}
\usepackage{bm}
\usepackage{amsmath,amssymb,amsfonts,bbold}
\usepackage[normalem]{ulem}
\usepackage{setspace}
\usepackage{nicefrac}

\newcommand{\beq}{\begin{equation}}
\newcommand{\eeq}{\end{equation}}
\newcommand{\beqa}{\begin{eqnarray}}
\newcommand{\eeqa}{\end{eqnarray}}

\newcommand{\MnO}{MnO$_2$}
\newcommand{\MnOO}{MnO$_3$}
\newcommand{\threebyone}{$(3 \times 1)$}
\newcommand{\sixbytwo}{$c(6 \times 2)$}

\graphicspath{{fig/}}

\begin{document}

\title{Reversible Tuning of Collinear versus Chiral Magnetic Order by Chemical Stimulus}

\author{Jing Qi}
		\affiliation{Physikalisches Institut, Experimentelle Physik II, Universit\"{a}t W\"{u}rzburg, Am Hubland, 97074 W\"{u}rzburg, Germany}
\author{Paula M. Weber}
		\affiliation{Physikalisches Institut, Experimentelle Physik II, Universit\"{a}t W\"{u}rzburg, Am Hubland, 97074 W\"{u}rzburg, Germany}
\author{Tilman Ki{\ss}linger}
        \affiliation{Solid State Physics, Friedrich-Alexander-Universit\"{a}t Erlangen-N\"{u}rnberg, Staudtstrasse 7, 91058 Erlangen, Germany}
\author{Lutz Hammer} 
        \affiliation{Solid State Physics, Friedrich-Alexander-Universit\"{a}t Erlangen-N\"{u}rnberg, Staudtstrasse 7, 91058 Erlangen, Germany}
\author{M. Alexander Schneider}
        \email[corresponding author: \\]{alexander.schneider@fau.de}
        \affiliation{Solid State Physics, Friedrich-Alexander-Universit\"{a}t Erlangen-N\"{u}rnberg, Staudtstrasse 7, 91058 Erlangen, Germany}
\author{Matthias Bode}
        \email[corresponding author: \\]{bode@physik.uni-wuerzburg.de}
	    \address{Physikalisches Institut, Experimentelle Physik II, Universit\"{a}t W\"{u}rzburg, Am Hubland, 97074 W\"{u}rzburg, Germany}
	    \address{Wilhelm Conrad R{\"o}ntgen-Center for Complex Material Systems (RCCM), 
	    			Universit\"{a}t W\"{u}rzburg, Am Hubland, 97074 W\"{u}rzburg, Germany}
	
\date{\today}

\begin{abstract}
The Ruderman-Kittel-Kasuya-Yosida (RKKY) interaction mediates collinear magnetic interactions 
via the conduction electrons of a non-magnetic spacer, resulting in a ferro- or antiferromagnetic magnetization in magnetic multilayers.  
The resulting spin-polarized charge transport effects have found numerous applications. 
Recently it has been discovered that heavy non-magnetic spacers are able 
to mediate an indirect magnetic coupling that is non-collinear and chiral. 
This Dzyaloshinskii-Moriya-enhanced RKKY (DME-RKKY) interaction causes the emergence 
of a variety of interesting magnetic structures, such as skyrmions and spin spirals.  
Applications using these magnetic quasi-particles require a thorough understanding and fine-tuning of the balance 
between the Dzyaloshinskii-Moriya interaction and other magnetic interactions, e.g., the exchange interaction and magnetic anisotropy contributions. 
Here, we show by spin-polarized scanning tunneling microscopy that the spin structure of manganese oxide chains on Ir(001) 
can reproducibly be switched from chiral to collinear antiferromagnetic interchain interactions 
by increasing the oxidation state of MnO$_2$ while the reverse process can be induced by thermal reduction. 
The underlying structural change is revealed by low-energy electron diffraction intensity data (LEED-IV) analysis. 
Density functional theory calculations suggest that the magnetic transition may be caused 
by a significant increase of the Heisenberg exchange upon oxidation.
\end{abstract}

\maketitle

{\em Introduction ---}
The coupling between magnetic layers separated by non-magnetic spacers 
has attracted considerable interest over the past 40 years \cite{Fert1995,Stiles1999}.  
This interest was mainly driven by the discovery of spin-polarized charge transport effects, 
such as the giant magnetoresistance (GMR) effect which---depending on the relative alignment 
of the layer magnetization---leads to a substantial resistance change. 
It was soon realized that the interlayer exchange coupling is driven by the Ruderman-Kittel-Kasuya-Yosida 
(RKKY) coupling \cite{1954-PR-Indirect,1956-Prog.Theor.Phys.-A,PhysRev.106.893,1960-Phys.Rev.-Anisotropic}
that mediates magnetic interactions via the conduction electrons of a non-magnetic spacer \cite{Bruno1991}.  
 
Whereas the conventional RKKY interaction only results in collinear coupling terms, 
i.e., a ferromagnetic or antiferromagnetic arrangement of the layers, 
another long predicted \cite{FL1980} yet only recently discovered long-range interaction is able to mediate chiral magnetic coupling 
between magnetic chains \cite{2019-Nat.Commun.-Indirect,2019-PRB-Structural} or layers \cite{GCY2008,Han2019,Pacheco2019} 
in two-dimensional or three-dimensional host systems, respectively.  
Similar to conventional collinear RKKY, this novel interaction is also mediated 
by conduction electrons of the substrate which are polarized by the magnetic material.  
Due to the high spin-orbit coupling (SOC) of the spacer, however, a Dzya\-lo\-shinskii-Moriya (DM) type enhancement takes place,
resulting in an asymmetric indirect exchange interaction and chiral magnetic coupling.  
Therefore, this new interaction will be termed {\em DM-enhanced RKKY} (DME-RKKY) interaction \cite{2019-Nat.Commun.-Indirect} hereafter
in the spirit of the original proposal \cite{FL1980}, however, we would like to point out that the term {\em interlayer DMI} has also been used before \cite{Han2019,Pacheco2019}.  

The DME-RKKY interaction is of high fundamental and practical interest as it may ``provide the capability 
to further tailor topological spin textures, in not only one-dimensional (1D) or two-dimensional (2D) but also 3D space'' \cite{Dohi2022}.  
In particular, this would open up the possibility to create and utilize so-called skyrmions, 
which are extremely robust against thermal or quantum fluctuations \cite{Braun2012} at all spatial dimensions. 
These DMI-driven spin structures may potentially enable novel spintronic applications, 
such as racetrack memories \cite{Ummelen2017}, spin valves with heavy-metal cap layers \cite{DLL2021}, 
or even spin-orbit torque-based logic devices \cite{SCL2020}, all combining the advantages of robustness and low energy consumption. 

Whether topologically protected chiral spin structures form or not critically depends on the balance 
between DMI and other magnetic interactions, such as the exchange interaction and the magneto-crystalline anisotropy.  
Particularly fascinating and promising for applications are concepts 
where this balance can reversibly be fine-tuned by an external stimulus.
Various methods have been proposed, including tuning of the oxygen coverage \cite{Belabbes2016}, 
hydrogen adsorption \cite{Yang2020}, or the application of electric fields \cite{Yang2018}, 
but none of these concepts has yet been realized in the context of the DME-RKKY interaction.  

A chiral interchain coupling mediated by DME-RKKY interaction was initially found in MnO$_2$ chains grown on Ir(001) \cite{2019-Nat.Commun.-Indirect}. 
Here we present an extended spin-polarized scanning tunneling microscopy (SP-STM) study showing that
this interchain coupling can reversely be switched between chiral non-collinear and collinear antiferromagnetic (AFM) order
by tuning the oxygen partial pressure $p_{\rm O_{2}}$ in the final preparation step. 
Structural analysis of the energy-dependent intensity of low-energy electron diffraction beams (LEED-IV) reveals 
that these different spin configurations arise from a structural transition from MnO$_2$ at low $p_{\rm O_{2}}$ 
to \MnOO\ chains with the addition of another oxygen atom at the Ir bridge site, O$_{\textrm{br}}$, at high $p_{\rm O_{2}}$. 
Our work presents a new path toward ``spin-orbitronics''  \cite{Yang2020} 
by introducing a reversible and reproducible manipulation of the magnetic structure 
in indirectly coupled magnetic transition-metal oxide systems via chemically tuning their stoichiometry.
Moreover, our work provides an addition to the toolbox of tailoring topological spin textures by the DME-RKKY interaction.

\begin{figure*}
    \includegraphics[width=1\textwidth]{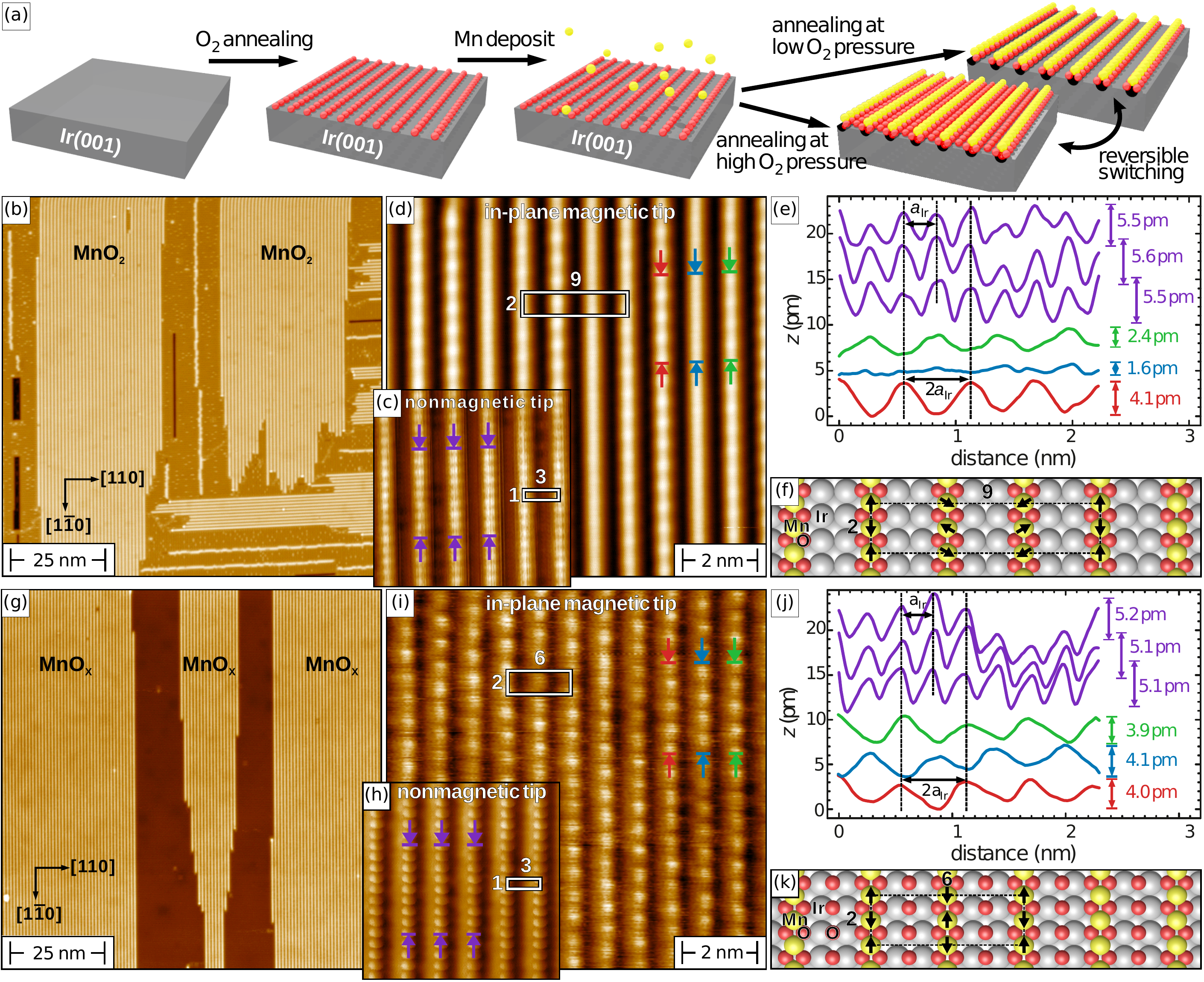}
	\caption{(a) Scheme of the sample preparation procedures. 
		Depending on the oxygen pressure $p_{\rm O_{2}}$ during the final annealing step, 
		dif\-fe\-rent MnO$_{x}$ stoichiometries with $x = 2$ or $x > 2$ are obtained. 
		(b) Large-scale STM topographic image of a sample prepared at low $p_{\rm O_{2}}$.  
		MnO$_{2}$ chains are oriented along the $[110]$ and $[1\overline{1}0]$ direction of Ir(001)
		(scan parameters: $U = 100$\,mV, $I = 500$\,pA). 
		(c),(d) Atomic resolution images scanned with a non-magnetic tungsten tip 
		and a magnetic Mn tip ($U = 100$\,mV, $I = 1$\,nA). 
		(e) Line profiles measured on the chains indicated by the correspondingly colored markers in (c) and (d) 
		showing a doubling of the $a_{\rm Ir}$ period in (d) as well as a systematic variation of amplitudes and phases 
		between green, blue, and red line profiles, characteristic for chiral magnetic order across MnO$_{2}$ chains. 
		(f) Model of the ($9 \times 2$) MnO$_{2}$ structure. Grey, red, and yellow spheres represent Ir, O, and Mn atoms, respectively. 
		Mn spin orientation is marked by black arrows. 
		(g) Large scale STM image of a similar surface as in (a), but prepared at high $p_{\rm O_{2}}$ ($U = 100$\,mV, $I = 50$\,pA). 
		(h),(i) Atomic resolution STM images of the \sixbytwo\ MnO$_{x}$ chain structure 
		scanned with a non-magnetic and a Mn-coated magnetic tip, 
		respectively [(h) $U = 50$\,mV, $I = 1$\,nA; (i) $U = 300$\,mV, $I = 0.5$\,nA]. 
		(j) Line profiles along the lines in (h) and (i), 
		showing a doubling of the periodicity in (i) compared to that of (h) and an anti-phase relation between neighboring rows. 
		(k) Model of the \sixbytwo\ MnO$_{x}$ structure. }
	\label{fig:Fig1}
\end{figure*}

{\em Sample preparation ---} The four steps of our sample preparation procedure are schematically represented in Fig.\,\ref{fig:Fig1}(a). 
It consists of (i) the preparation of a clean Ir(001), (ii) the formation of an oxygen-terminated Ir(001)--($2 \times 1$) adsorbate structure, 
(iii) sub-monolayer (ML) Mn deposition, and (iv) a final oxygen annealing step.  
We found that the oxygen pressure during this concluding step is decisive for the stoichiometry of the resulting MnO$_{x}$ chains.  
Whereas a high exposure of $\approx 13.5$\,Langmuir (L) at $p_{\rm O_{2}} \geqslant 1 \cdot 10^{-7}$\,mbar 
yields an oxygen-rich MnO$_{x > 2}$ structure, exposure to $\approx 1.3$\,L 
at $p_{\rm O_{2}} \leqslant 5 \cdot 10^{-8}$\,mbar gives the oxygen-poor MnO$_{2}$ structure. 
A careful discussion of the transferability of the preparation parameters between the different laboratories, 
SP-STM and LEED-IV data acquisition and analysis, 
as well as density functional theory (DFT) procedures is provided in the supplementary material \cite{SupplMat}.

{\em Experimental Results (STM) ---}
Figure \ref{fig:Fig1}(b)-(d) present spin-averaged and SP-STM results 
obtained on MnO$_{2}$ chains prepared at low $p_{\rm O_{2}}$.  
Fig.\,\ref{fig:Fig1}(b) shows a typical overview STM topographic image of a sample partially covered by MnO$_{2}$ chains. 
The chains are highly ordered, periodic, and oriented along the Ir $[110]$ and $[1\overline{1}0]$ directions. 
Previous studies showed that the Mn atoms reside above a missing-row structure 
with only indirect interaction with the Ir substrate via oxygen atoms. 
Each Mn atom is coordinated to four surrounding oxygen atoms, which in turn bind to 
one substrate Ir atom and two Mn atoms \cite{2016-PRL-Self-Organized}. 
Fig.\,\ref{fig:Fig1}(c) shows an atomically resolved STM image of the MnO$_2$ chains scanned with a non-magnetic tungsten tip.
We recognize a ($3 \times 1$) unit cell (white rectangle), in good agreement 
with published data \cite{2016-PRL-Self-Organized,2019-Nat.Commun.-Indirect,2019-PRB-Structural}. 
Line profiles along three adjacent MnO$_{2}$ chains marked by purple markers in Fig.\,\ref{fig:Fig1}(c) 
are presented in the upper part of Fig.\,\ref{fig:Fig1}(e). 
The periodicity is consistent with the Ir lattice constant $a_{\rm Ir}\approx 2.71\,$\AA.

Figure~\ref{fig:Fig1}(d) shows a corresponding data set, but scan\-ned with a spin-sensitive tip. 
In good agreement with Ref.\,\onlinecite{2019-Nat.Commun.-Indirect} 
we recognize a magnetic ($9 \times 2$) unit cell (white rectangle).
Line profiles taken along three adjacent MnO$_{2}$ chains reveal 
that the periodicity has doubled to $2a_{\rm Ir}$, see lower part of Fig.\,\ref{fig:Fig1}(e). 
The corrugation amplitude along the chains systematically varies between $1.6$\,pm and $4.1$\,pm 
and the line profiles are phase-shifted to one another.  
As sketched in Fig.\,\ref{fig:Fig1}(f), this modulation of the magnetic contrast 
can consistently be explained \cite{2019-Nat.Commun.-Indirect} by a magnetic ($9 \times 2$) unit cell caused by collinear AFM order along the chains and a chiral $120^{\circ}$ spin spiral perpendicular to the chains.    

Figure~\ref{fig:Fig1}(g) shows a typical large-scale STM image of a sample prepared similar to the one 
presented in Fig.\,\ref{fig:Fig1}(b), but with a final annealing step performed at high $p_{\rm O_{2}}$.  
We again find chains which are highly ordered, periodic, and oriented along the Ir $[110]$ and $[1\overline{1}0]$ directions.
Since the final annealing step ($T_{\rm ann} = 1020$\,K) was performed at high $p_{\rm O_{2}}$, 
and since the \MnO\ chains only decompose in vacuum at temperatures above 1070\,K, 
we speculate that the oxygen excess results in the formation of MnO$_{x > 2}$. 
This assumption will be confirmed below by LEED-IV structural analysis.
Fig.\,\ref{fig:Fig1}(h) shows atomically resolved data of MnO$_{x > 2}$ chains scanned with a non-magnetic tip. 
We recognize a ($3 \times 1$) unit cell (white rectangle), i.e., the same as in Fig.\,\ref{fig:Fig1}(c).  
Line profiles along three adjacent MnO$_{x > 2}$ chains 
indicated by purple markers in Fig.\,\ref{fig:Fig1}(h) are shown in the upper part of Fig.\,\ref{fig:Fig1}(j).  
Again, the periodicity agrees with $a_{\rm Ir}$.
The striking resemblance of Fig.\,\ref{fig:Fig1}(h) and Fig.\,\ref{fig:Fig1}(c) implies that the Mn core of the oxide chain remains unchanged.

SP-STM data of the MnO$_{x > 2}$ chain structure revealing a \sixbytwo\ unit cell (white rectangle) are presented in Fig.\,\ref{fig:Fig1}(i). 
Line profiles along three adjacent MnO$_{x > 2}$ chains
marked in Fig.\,\ref{fig:Fig1}(i) are shown in the lower part of Fig.\,\ref{fig:Fig1}(j). 
A $2a_{\rm Ir}$ periodicity and a $\pi$ phase shift between adjacent chains are immediately apparent. 
Furthermore, unlike the MnO$_2$ chain structure in Fig.\,\ref{fig:Fig1}(d), the corrugation of $(4.0 \pm 0.1)$\,pm 
measured along the MnO$_{x > 2}$ chains remains the same for adjacent chains. 
These observations imply a spin structure which is not only collinear AFM along the chains, but also collinear across adjacent chains.  
Fig.\,\ref{fig:Fig1}(k) shows this presumed spin structure of MnO$_{x > 2}$ 
with its \sixbytwo\ magnetic unit cell 
	\footnote{A series of control experiments was performed to exclude that (i) the oxygen pressure applied in the second step 
	or (ii) the amount of Mn deposited onto the Ir substrate plays a critical role 
	for the formation of the \sixbytwo\ and the ($9 \times 2$) magnetic unit cell.  
	These studies led to the conclusion that only the oxygen pressure during the last oxygen annealing step 
	(step four forming MnO$_{x}$ chains) determines the resulting MnO$_{x}$ structure, 
	while the other parameters do not matter in a general sense. }.

Importantly, we were able to reversibly and repeatedly switch between the collinear AFM spin structure 
of \sixbytwo\ MnO$_{x > 2}$ and the non-collinear, chiral ($9 \times 2$) MnO$_{2}$ chains 
by annealing at $T_{\rm ann} = 1020$\,K in high or low $p_{\rm O_{2}}$, respectively 
(see Ref.\,\onlinecite{SupplMat} for exemplary results).
We would like to highlight that this transition does not require any further Mn supply, 
indicating that only the oxidation state changes whereas the Mn remains on the Ir(001) surface, 
without any desorption into the gas phase or diffusion into the substrate.  

\medskip
\begin{figure}
	\includegraphics[width=\columnwidth]{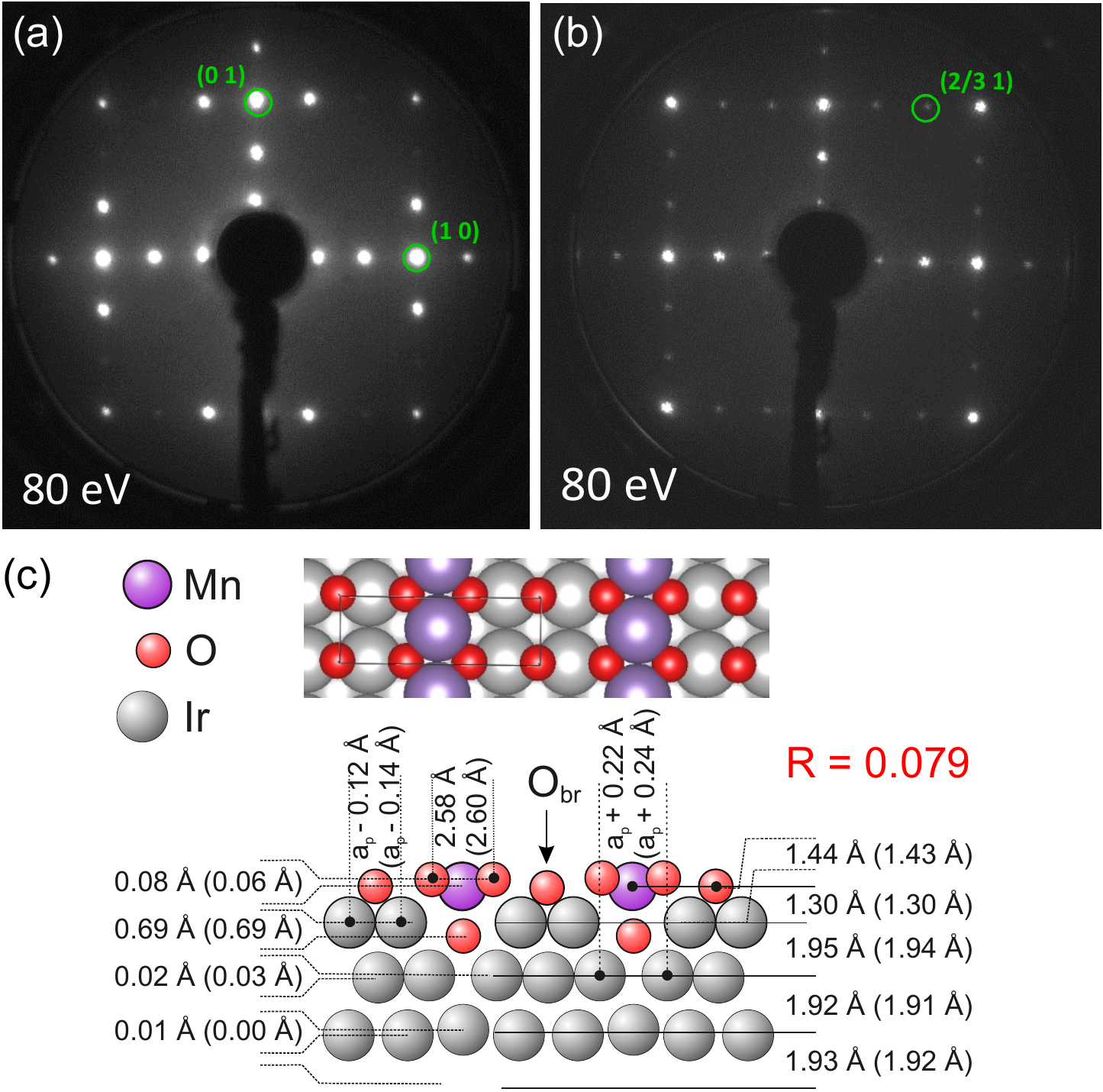}
	\caption{LEED patterns of $1/3$\,ML Mn oxidized in (a) $p_{\rm O_{2}} = 5 \cdot 10^{-8}$\,mbar molecular O$_2$ 
	and (b) in $p_{\rm NO_2} = 1 \cdot 10^{-6}$\,mbar (exact conditions see text). 
	The intensity spectrum of the spot marked in (b) is shown in Fig.\,\ref{fig:Fig3}. 
	(c) Top and side view of the $(3 \times 1)$ MnO$_3$+O$_{\textrm{br}}$ structure 
	and relevant structural parameters as obtained by LEED-IV structural analysis of the preparation in (b). 
	The resulting parameters of the DFT energy minimization are given in brackets. }  
	\label{fig:Fig2}
\end{figure}
{\em Experimental Results (LEED-IV) ---}
It is evident that the two distinct magnetic unit cells observed for MnO$_x$ chains 
prepared by high and low $p_{\rm O_{2}}$ must have a structural reason. 
Based on our previous work on the CoO$_x$ chains on Ir(001) \cite{2017-PRB-Monatomic}, 
we expected that \MnO\ could also be further oxidized to \MnOO.
However, for Co this was only achieved with an oxidizing agent (NO$_2$) that provides O atoms easily \cite{2017-PRB-Monatomic}.
We therefore first checked if \MnOO\ chains could be produced from \MnO\ by oxidation with NO$_2$.
To obtain the crystallographic structure from LEED-IV analysis with high quality and low error margins, 
we deposited 1/3 ML Mn/Ir(001) which leads to a surface homogeneously covered 
with \MnO\ chains in a $(3 \times 1)$ superstructure after oxidation in $p_{\rm O_{2}} = 5 \cdot 10^{-8}$\,mbar.
A LEED pattern of this surface is presented in Fig.\,\ref{fig:Fig2}(a).  
Comparison to LEED spectra obtained previously \cite{2016-PRL-Self-Organized} 
confirms that the chains have the expected \MnO\ structure.   

\begin{figure}
	\includegraphics[width=0.85\columnwidth]{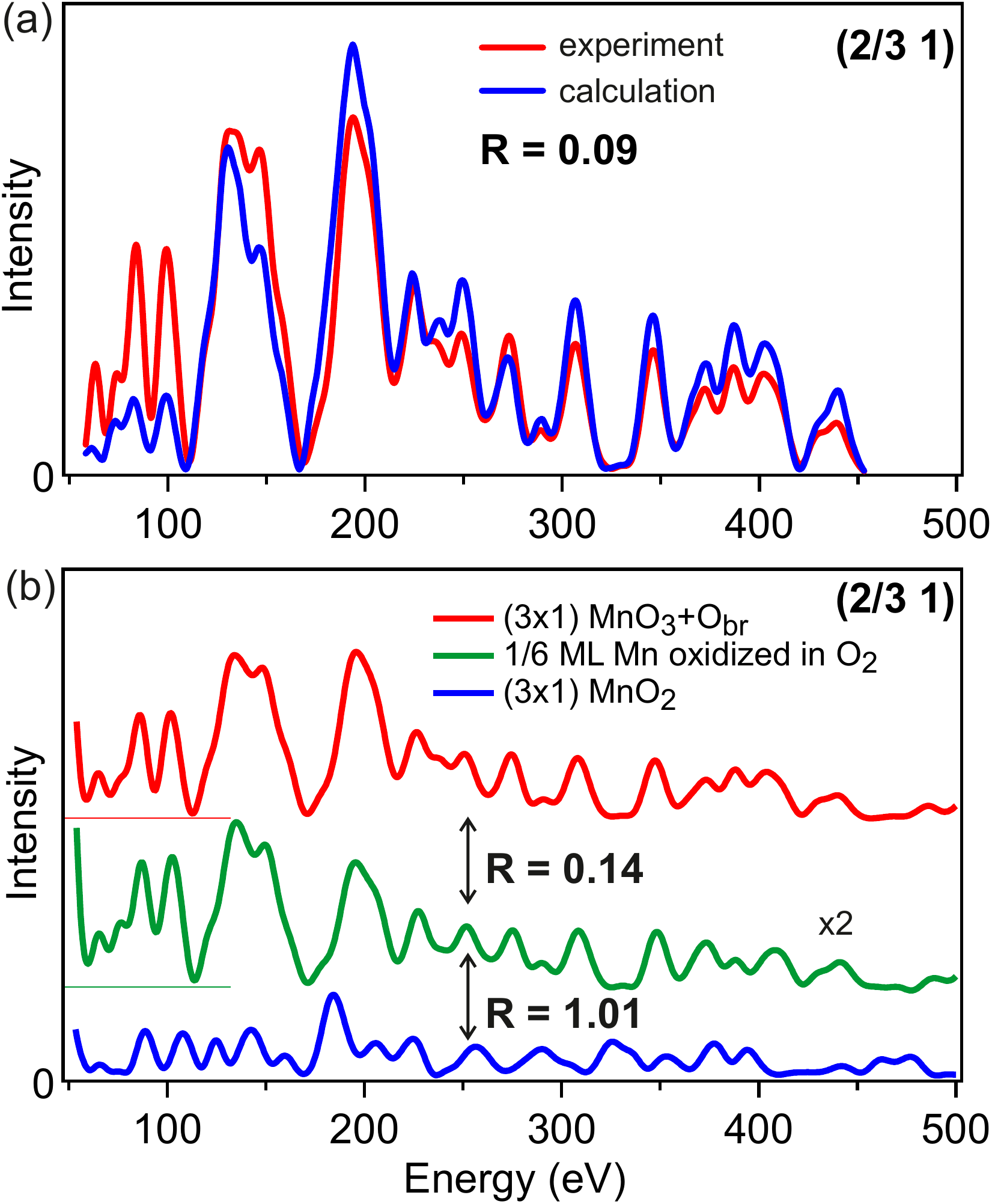}
	\caption{Intensity spectra of the k$_{||}=(\nicefrac{2}{3}\ 1)$ beam. 
	(a) Comparison of the normalized intensity between the experimental (red) 
	and the calculated (blue) spectrum of the \MnOO+O$_{\textrm{br}}$ structure (single beam $R$ factor $R = 0.091$). 
	(b) Comparison between experimental spectra obtained from 
	the \threebyone MnO$_2$ at \nicefrac{1}{3} ML Mn(lowest curve, blue), of the \threebyone \MnOO+O$_{\textrm{br}}$ 
	prepared by oxidation in NO$_2$ (top curve, red), 
	and a preparation comparable to that when the \sixbytwo\ magnetic super cell occurs (middle curve, green). 
	The middle curve is scaled by a factor of 2. 
	The visual agreement between the middle and upper curve is substantiated by an $R$-factor of $R = 0.14$. 
	Curves in (b) are offset for clarity. } 
	\label{fig:Fig3}
\end{figure}

Subsequently, we oxidized the surface in $p_{\rm NO_2} = 10^{-6}$ mbar at 870\,K 
and performed two annealing steps at 530\,K. 
The final cool-down was carried out without NO$_2$ flux to ensure the desorption of remaining NO$_x$ from the surface.
The LEED image of the resulting surface is shown in Fig.\,\ref{fig:Fig2}(b).  
Again, a \threebyone\ LEED pattern is observed with, however, very different spot intensities. This unambiguously proves a changed surface structure.
A LEED-IV analysis with a Pendry $R$-factor of $R= 0.083$ reveals that the structure obtained after NO$_2$ oxidation
is indeed the expected \MnOO\ chain structure (analogous to CoO$_3$ \cite{2017-PRB-Monatomic}), 
only with the important difference that the Ir bridge sites between the chains are all occupied by additional oxygen atoms (with error margin of 17\%). 
The model was also tested in a structural relaxation calculation using DFT.  
All structural parameters determined experimentally agree within the single-digit picometer range to the calculated values after scaling the DFT $a_\textrm{Ir}$ lattice parameter to the experimental one (Fig.\,\ref{fig:Fig2}(c)). 
An example of the excellent agreement between measured and calculated LEED spectra is shown in Fig.\,\ref{fig:Fig3}(a). 
For the complete data set see Ref.\,\onlinecite{SupplMat}. 

For an Ir(001) surface fully covered with \MnO\ chains, further oxidation in O$_2$ 
up to $p_{\rm O_{2}} = 10^{-6}\,\textrm{mbar}$ was not possible, i.e., the chains remained in their \MnO\ state. 
Only for partially covered surfaces, e.g., at 1/6\,ML Mn, further oxidation became possible. 
We suspect that partially covered Ir(001) provides (defect) sites for O$_2$ dissociation. 
However, the reduced order also complicates the LEED-IV analysis.
The proof of oxidation becomes possible when looking at the third-order diffraction spots 
that do not contain intensity contributions from the coexisting ($2\times 1$)O phase.
This is shown in Fig.\,\ref{fig:Fig3}(b) for the $k_{||} = (\nicefrac{2}{3}, 1)$ beam.
Visual inspection of the spectra already reveals that those of the highly oxidized phase at 1/6\,ML Mn 
belong to the pure (1/3 ML) \MnOO+O$_\textrm{br}$ rather than to the \MnO\ phase. 
An $R$ factor analysis between a set of experimental third-order beams of the 1/6\,ML phase 
and the \MnOO+O$_\textrm{br}$ structure yields $R = 0.12$, whereas comparison to the \MnO\ structure 
gives $R = 0.81$ \cite{SupplMat}. 
We can therefore be confident that the preparation with 1/6\,ML Mn oxidized at $p_{\rm O_{2}} = 10^{-6}\,\textrm{mbar}$  
results in the \MnOO+O$_\textrm{br}$ phase and that this structure 
gives rise to the \sixbytwo\ magnetic unit cell observed in Fig.\,\ref{fig:Fig1}(g)-(i). 
Furthermore, a phase stability analysis by DFT \cite{2017-PRB-Monatomic, Reuter2001, Rogal2007} supports the view 
that only the \MnOO+O$_\textrm{br}$ and \MnO\ chains appear as stable phases \cite{SupplMat}.

\medskip
{\em Discussion ---}
The SP-STM results presented in Fig.\,\ref{fig:Fig1} show that the magnetic order across adjacent MnO$_{x}$ chains 
can controllably and reversibly be switched between a collinear AFM and non-collinear chiral state 
by annealing at high or low $p_{\rm O_{2}}$, respectively. 
Structural LEED-IV analysis reveals that this different interchain coupling comes along with a different surface stoichiometry.  
While samples annealed a low $p_{\rm O_{2}}$ exhibit MnO$_{2}$ chains, 
annealing at high $p_{\rm O_{2}}$ results in a higher oxidation state, 
where a third O atom is relaxed in the vacancy row underneath the Mn atom
and another O occupies the Ir bridge site midway between the MnO$_{3}$ chains.   

The chiral $120^{\circ}$ magnetic coupling observed between MnO$_{2}$ chains across the intermediate Ir(001) substrate 
was explained by the DME-RKKY interaction \cite{2019-Nat.Commun.-Indirect}.  
Similar to the well-established, collinear RKKY interaction, the DME-RKKY interaction is mediated 
by conduction electrons of the substrate which are polarized by the magnetic chains.  
Due to the high spin-orbit coupling of Ir, this indirect exchange becomes asymmetric, resulting in a chiral magnetic coupling.  
We have to bear in mind, however, that a chiral spin structure can only develop if the DMI is able 
to overcome the Heisenberg exchange interaction $J$ and the anisotropy $K$ \cite{2007-Nat.-Chiral,Schwefling2016}. 
In a slightly simplified picture the transition occurs at $D^2 > 4JK$ \cite{2007-Nat.-Chiral}.   
Since $J$ is extremely small for MnO$_2$/Ir(001), about 1\,meV per Mn atom only, 
the DMI can trigger the spin spiral state observed in Fig.\,\ref{fig:Fig1}(d) \cite{2016-PRL-Self-Organized,2019-Nat.Commun.-Indirect}.  

To get an idea of the relevant $J$ values which drive the transition to the collinear state for TMO chains
with the stoichiometry MnO$_{x > 2}$, we calculated the magnetic intra- and interchain coupling 
for \MnO, \MnOO, a hypothetical \MnO +O$_{br}$, and the \MnOO+O$_\textrm{br}$ structures using spin-resolved, collinear DFT+U.
All structures show a strong AFM intrachain coupling of 20 to 30\,meV per Mn atom, 
in agreement with our experimental findings presented in Fig.\,\ref{fig:Fig1}(d,i). 
The interchain coupling quantitatively depends on the amount of oxygen in the surface structure.
As reported before, we calculate a weak AFM interchain coupling of $\approx 1.5$\,meV per atom 
for the \MnO\ chains \cite{2016-PRL-Self-Organized}. 
This weak AFM interchain coupling essentially vanishes for the hypothetical \MnO +O$_{\rm br}$ structure. 
For the \MnOO\ structure we find a much stronger AFM interchain coupling of $7\,\textrm{meV}$  per atom, 
and eventually a FM interchain coupling of  $5\,\textrm{meV}$ when the O$_{br}$ is added to the structure.
Despite the fact that in the DFT approximation a $(3 \times 2)$ magnetic cell for the experimentally found \MnOO +O$_{\rm br}$ is predicted, 
the results clearly show that indirect coupling via the substrate is substantially increased by the additional oxygen atom underneath the Mn atom. 
The O$_{\rm br}$ adds a second interference path for the RKKY interaction and may change size and sign of the interchain interaction.
To resolve the discrepancy with respect to the experimentally observed magnetic cells, 
we suggest to use our robust structural analysis as a basis for more refined calculations 
which also consider non-collinear magnetic order and spin-orbit coupling. 

\begin{acknowledgments}
{\em Acknowledgement ---}This work was supported by the Deutsche Forschungsgemeinschaft 
(DFG, German Research Foundation) under Germany's Excellence Strategy through the W{\"u}rzburg-Dresden 
Cluster of Excellence on Complexity and Topology in Quantum Matter (ct.qmat) (EXC 2147, project-id 390858490).
\end{acknowledgments}

\bibliography{Bibliography_v23}

\end{document}


\title{Supplemental Material for \\ ``Reversible Tuning of Collinear versus Chiral Magnetic Order \\ by Chemical Stimulus"}

\author{Jing Qi}
		\affiliation{Physikalisches Institut, Experimentelle Physik II, Universit\"{a}t W\"{u}rzburg, Am Hubland, 97074 W\"{u}rzburg, Germany}
\author{Paula M. Weber}
		\affiliation{Physikalisches Institut, Experimentelle Physik II, Universit\"{a}t W\"{u}rzburg, Am Hubland, 97074 W\"{u}rzburg, Germany}
\author{Tilman Ki{\ss}linger}
        \affiliation{Solid State Physics, Friedrich-Alexander-Universit\"{a}t Erlangen-N\"{u}rnberg, Staudtstrasse 7, 91058 Erlangen, Germany}
\author{Lutz~Hammer} 
        \affiliation{Solid State Physics, Friedrich-Alexander-Universit\"{a}t Erlangen-N\"{u}rnberg, Staudtstrasse 7, 91058 Erlangen, Germany}
\author{M. Alexander Schneider}
        \email[corresponding author: \\]{alexander.schneider@fau.de}
        \affiliation{Solid State Physics, Friedrich-Alexander-Universit\"{a}t Erlangen-N\"{u}rnberg, Staudtstrasse 7, 91058 Erlangen, Germany}
\author{Matthias Bode}
        \email[corresponding author: \\]{bode@physik.uni-wuerzburg.de}
	    \address{Physikalisches Institut, Experimentelle Physik II, Universit\"{a}t W\"{u}rzburg, Am Hubland, 97074 W\"{u}rzburg, Germany}
	    \address{Wilhelm Conrad R{\"o}ntgen-Center for Complex Material Systems (RCCM), 
	    Universit\"{a}t W\"{u}rzburg, Am Hubland, 97074 W\"{u}rzburg, Germany}
	
\date{\today}

\maketitle
	 
\section*{I. Experimental Details}

\subsection*{Sample preparation procedures}
Since the samples were prepared at two laboratories in different ultra-high vacuum (UHV) systems, 
the values of temperature and particularly of the oxygen partial pressure have to be taken as guide only. 
Preparation temperatures were determined by a pyrometer (Ircon Ultimax UX-20P, operated at an emissivity $\epsilon = 0.33$) 
in W{\"u}rzburg (JMU) and a type-K thermocouple in Erlangen (FAU), respectively.  
We estimate the error margins to $\pm 50$\,K at JMU and $\pm 20$\,K at FAU.  

In both laboratories molecular oxygen gas with a nominal purity of 99.999\%\ was used for surface oxidation.  
In the FAU laboratory the UHV chamber was back-filled with oxygen via a leak valve 
which is positioned a few tens of centimeters away from both, the sample and the vacuum gauge, and not facing the sample.  
Therefore, it can reasonably be assumed that the measured pressure, $p^{\rm mea}_{\rm O_{2}}$, 
closely matches the true oxygen pressure at the sample location. 
In contrast, at JMU the oxygen gas was dosed by a nozzle.
Since this nozzle is positioned just a few centimeters away from the sample surface, 
the local pressure will significantly exceed the readings of the pressure gauge. 
We estimate that the partial pressure at the sample location, $p^{\rm est}_{\rm O_{2}}$, 
is approximately one order of magnitude higher than the gauge readings in W{\"u}rzburg. 
Therefore, the pressure $p_{\rm O_{2}}$ indicated throughout the main paper corresponds 
to $p^{\rm mea}_{\rm O_{2}}$ for LEED-IV experiments performed at FAU 
and to $p^{\rm est}_{\rm O_{2}}$ for the (SP-)STM experiments performed at JMU.  

The Ir(100) surface was cleaned by repeated cycles of Ar$^+$ ion-sputtering 
followed by annealing to $T_{\rm ann} = 1500$\,K in oxygen atmosphere at $p_{\rm O_{2}} \approx 1 \cdot 10^{-8}$\,mbar. 
Subsequent sample annealing without oxygen yields a ($5 \times 1$) reconstruction 
characteristic for clean Ir(100) \cite{1976-Surf.Sci.-Preparation,1998-JCP-Surface,Schmidt2002}. 
Second, Ir is oxidized by annealing at $p_{\rm O_{2}} \approx 1 \cdot 10^{-8}$\,mbar ($T_{\rm ann} = 850$\,K) resulting in 
an oxygen-terminated Ir(100)--($2 \times 1$) adsorbate structure \cite{2016-PRB-Structure, 2000-JCP-Unusual}. 
Third, a sub-monolayer (ML) Mn amount equivalent to 0.16\,ML\,$\le \Theta \le 0.33$\,ML is deposited onto this surface at room temperature (RT). 
Finally, in step four, the preparation is completed by another oxygen annealing step at 1020\,K. 
\vfill \pagebreak

\subsection*{Methods}
Spin-polarized scanning tunneling microscopy (SP-STM) experiments were performed 
in a home-built UHV low-temperature scanning tunneling microscope (LT-STM) 
operated at a temperature $T = 4.5$\,K and a base pressure $p \leqslant 2.0 \cdot 10^{-10}$\,mbar. 
Electro-chemically etched polycrystalline tungsten (W) tips were introduced into the UHV chamber 
and flashed \textit{in vacuo} by electron bombardment at $\approx 1400$\,K. 
Spin-sensitive Mn-coated W tips were prepared by poking the STM tip by $\approx$ 1\,nm into MnO$_{x}$ chains on Ir(100).
We speculate that this procedure leads to the transfer of some magnetic material (Mn or MnO$_x$) from the sample surface onto the tip apex. 
Demagnetizing the spin-sensitive tip in general requires tip pulsing at $\approx -10$\,V (sample bias).

The LEED-IV investigation and corresponding preparations were carried out 
in an UHV chamber at FAU. 
The chamber was operated at a pressure in the low $10^{-10}$\,mbar regime and is equipped with standard surface preparation equipment for sputter cleaning and annealing samples, 
with an ErLEED optics (SPECS GmbH) and a room-temperature STM (RHK) allowing for real-space control of the quality of the surface preparation.
LEED intensity measurements were performed at normal incidence of the primary electron beam.

Spin-resolved DFT calculations were performed with the Vienna Ab-initio Simulation Package (VASP)\,\cite{vasp1,vasp2}, 
using projector augmented-wave (PAW) potentials\,\cite{paw} and the Perdew--Burke--Ernzerhof (PBE) 
exchange-correlation functional \cite{pbe} combined with DFT+U corrections \cite{ldaU} with a value of $U - J = 1.5$\,eV. 
Only collinear magnetism was considered. 
To capture the anti-ferromagnetic properties for all calculations a 7-layer Ir(100) $(6 \times 2)$ slab 
with 15\,\AA\ vacuum gap was set up and a $(3 \times 9 \times 1)$ Monkhurst-Pack $k$-point mesh was used.
\newpage

\subsection*{STM data of reversible switching between the two structures}
\begin{figure*}[h]
	\includegraphics[width=0.98\textwidth]{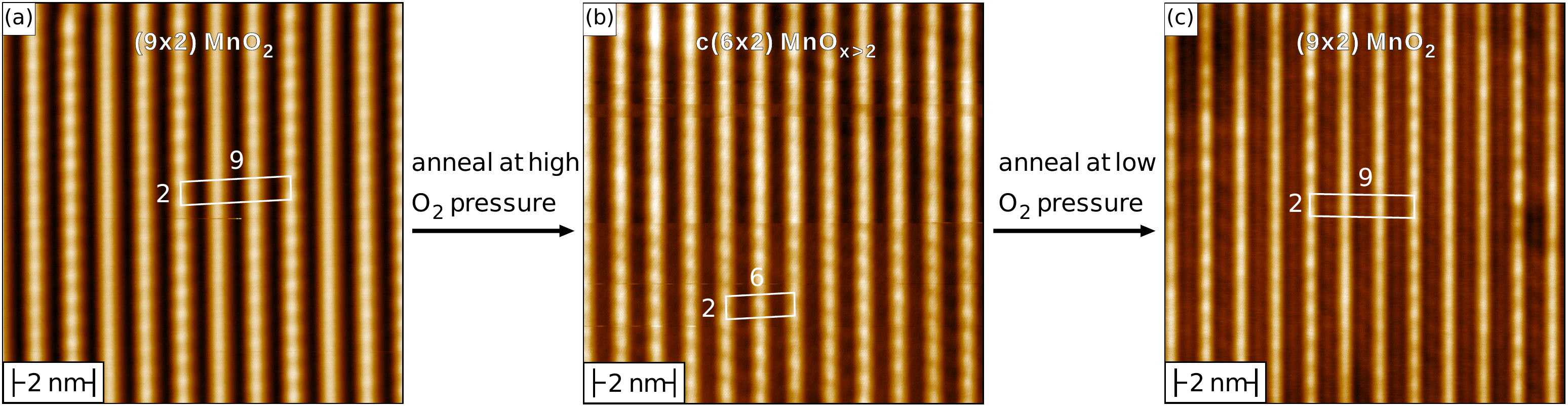}
	\caption{STM topographic images showing the exemplary consecutive switching between the collinear AFM spin structure 
			of $c(6 \times 2)$ MnO$_{x > 2}$ and the non-collinear, chiral ($9 \times 2$) MnO$_{2}$ chains 
			by annealing at $T_{\rm ann} = 1020$\,K at high or low $p_{\rm O_{2}}$. 
			White rectangles indicate the magnetic unit cells for each image.
			Scan parameters: (a) $U = 100$\,mV, $I = 1$\,nA; (b) $U = 10$\,mV, $I = 5$\,nA; 
			(c) $U = 10$\,mV, $I = 7$\,nA.} \label{fig:switch}
\end{figure*}

As shown in Fig.\,\ref{fig:switch}, we observed the reversible and reproducible switching between the collinear AFM spin structure 
of $c(6 \times 2)$ MnO$_{x > 2}$ and the non-collinear, chiral ($9 \times 2$) MnO$_{2}$ chains 
by annealing at $T_{\rm ann} = 1020$\,K in high or low $p_{\rm O_{2}}$, respectively.
Moreover, Mn supply is not necessary for this reversible switch to happen, indicating 
that only the oxidation state changes whereas the Mn remains on the Ir(001) surface, 
without any desorption into the gas phase or diffusion into the substrate.

\newpage

\section*{II. Structural analysis of M\lowercase{n}O$_3$+O$_\textrm{br}$}
\begin{figure*}[t]
	\includegraphics[width=0.88\textwidth]{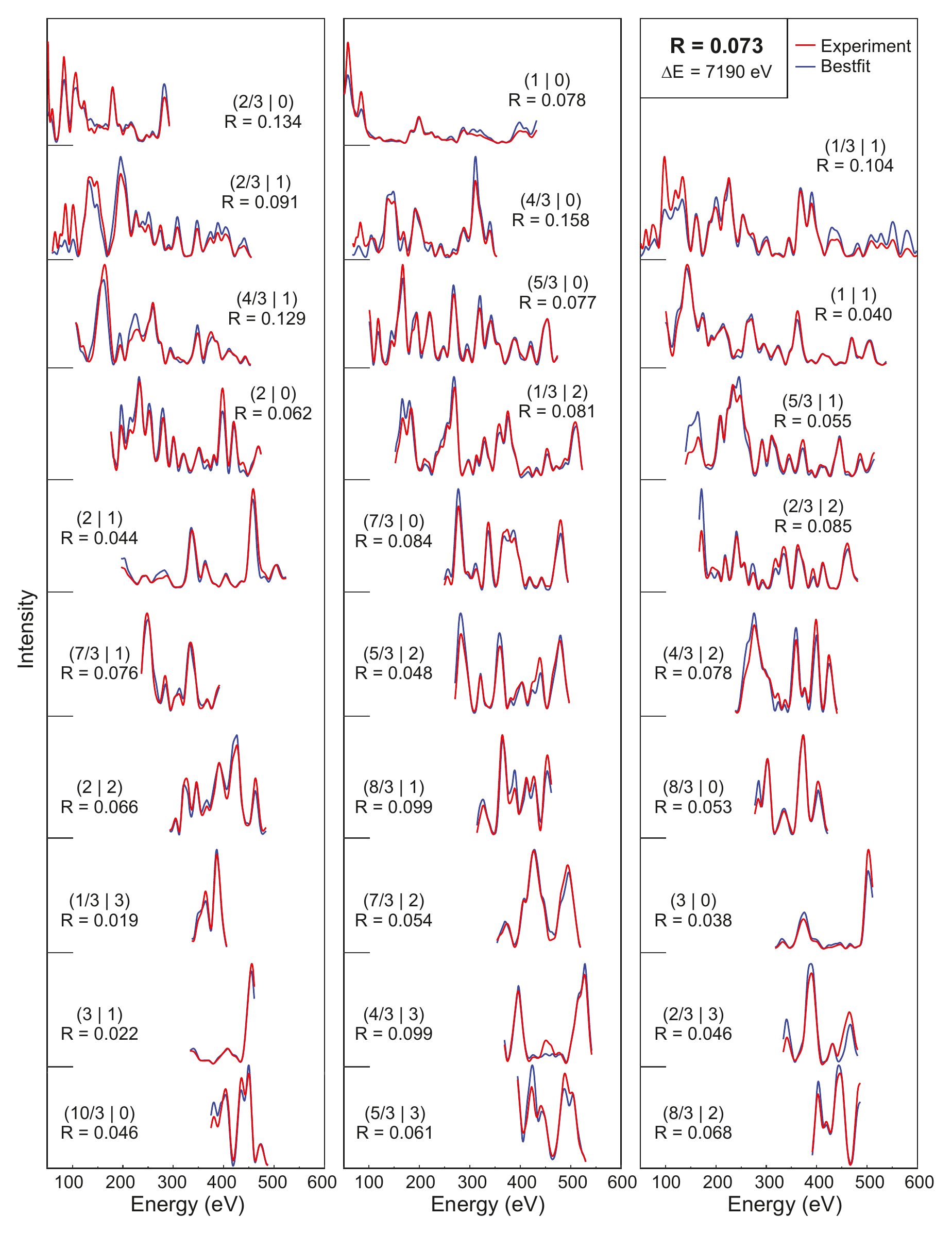}
	\caption{Compilation of all experimental LEED-IV spectra (red) entering the structural analysis together with their calculated counterparts (blue). In each panel the $R$-factor for that particular pair is given. For better visibility all spectra were normalized to the same intensity level.} \label{fig:bestfit}
\end{figure*}

\vspace{-0.4cm}
For the perfectly prepared Ir(100)-3$\times$1-MnO$_3$+O$_\textrm{br}$ phase the LEED pattern 
was recorded in the energy range 50 $\sim$ 600\,eV in 0.5\,eV steps and stacked into a video. 
By means of the new ViPErLEED-System \cite{Riva2022} the IV-spectra of all available beams with sufficient signal-to-noise ratio 
were automatically extracted, averaged over symmetrically equivalent beams, and moderately smoothed where necessary. 
This procedure finally resulted in a data base consisting of 29 inequivalent beams (7 integer and 22 fractional order) 
with a cumulated energy width of $\Delta E =$ 7190\,eV (1944\,eV / 5246\,eV), see Fig.\,\ref{fig:bestfit}.

The full-dynamically calculation of model intensities as well as the structure optimization, 
guided by the Pendry $R$-factor \cite{1980-J.Phys.C-Reliability} was performed also by using the ViPErLEED package \cite{Riva2022}, 
which manages a modified TensErLEED program code \cite{Blum2001}. 
The respective spherical-symmetric phase shifts and the energy-dependent part of the inner potential V$_{0i}$ 
were also calculated by the ViPErLEED package using Rundgren's code \cite{Rundgren2003}. 
Due to STM pre-informations we restricted the model search to models with MnO$_2$ chains with additional oxygen 
of variable amount below the Mn atom and between the chains alternatively in central hollow or bridge sites. 
Also, $pmm$-symmetry was assumed throughout. 
It turned out that leaving out any of these two oxygen species from the model, no $R$-factor value below $R = 0.20$ could be achieved. 
The same holds for putting oxygen into hollow sites between the manganese oxide chains. 
In contrast, for the model with MnO$_3$ chains and bridge-bonded oxygen (O$_\text{br}$) 
between the $R$-factor immediately fell below $R = 0.15$ in the first coarse fit cycle. 
We then restricted the further refinement to this model only.

In the final fit we adjusted all symmetry-allowed atomic displacements down to the 5$^{th}$ layer 
(4$^{th}$ layer for lateral parameters), which were 18 independent fit parameters in total. 
Additionally, we also varied the vibrational amplitudes of the three different oxygen species, 
the Mn atom and the 1$^{st}$ layer Ir atoms as well as the concentration of the low lying and bridge bonded oxygen atoms. 
The structural search was performed via TensorLEED \cite{Rous1986} on a grid with step width as fine as 0.2\,pm 
for all geometrical and vibrational parameters (concentration in 2\% steps) and subsequently verified by another full-dynamical calculation.  
In the course of the analysis, we also adjusted the optical potential as $V_{0i} = 5.05$\,eV, 
an offset for the inner potential $V_{00} = -0.5$\,eV and an effective angle of incidence $\Theta_{\rm eff} = 0.51^\circ$, 
which allowed, after averaging over all symmetry-equivalent beams, to simulate the slightly convergent incident electron beam in experiment.

\begin{figure}[h]
	\begin{minipage}[t]{0.38\textwidth} 
		\includegraphics[width=\columnwidth]{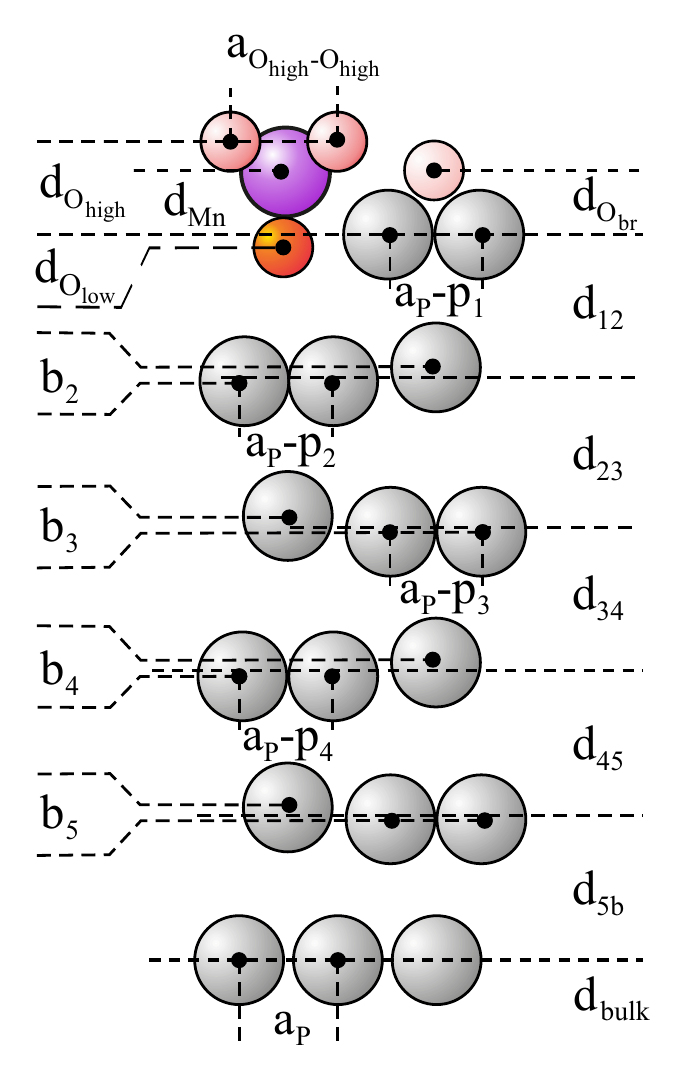}
	\end{minipage}
	\hfill
	\begin{minipage}[b]{0.5\textwidth}
		\caption{Definition of the geometrical parameters for the Ir(100)-3$\times$1-MnO$_3$+O$_\textrm{br}$ phase: 
		total buckling (b$_\text{i}$) and pairing amplitudes ($p_i$) within layer \emph{i}, 
		average layer spacings ($d_{i,i+1}$), vertical distances $d$ of Mn and oxygen species, 
		and lateral distances for the O$_{\text{high}}$ row separation (a$_\text{O$_{\text{high}}$-O$_{\text{high}}$}$).
		The bulk lattice parameters ($a_\text{p}$) of Ir is 2.7116\,{\AA} \cite{Arblaster2010} and the bulk layer distance $d_\textrm{bulk}$ is 1.9174\,\AA, accordingly.} 
		\label{fig:pardef} \vspace{2cm}
	\end{minipage}	
\end{figure}

The final fit achieved a Pendry $R$-factor as low as $R = 0.079$, 
which is among the best values ever achieved for a LEED-IV structure determination. 
A compilation of experimental and corresponding best-fit IV-spectra is displayed in Fig.\,\ref{fig:bestfit} together with single beam $R$-factors. 
Due to the large data base used for the analysis the total of $p = 28$ optimized parameter values 
could be determined with a still very high redundancy of $\rho = \Delta E / (4V_{0i} \cdot p) = 12.7$, i.e., they are on safe grounds. 
A separate file with coordinates of the best-fit structure is attached to this supplement. 
For a more intuitive representation of the structure, the atomic positions were transformed into quantities like layer spacings 
as well as atomic buckling and pairing amplitudes, which are defined in Fig.\,\ref{fig:pardef}. 
Tab.\,\ref{TableMnO3} summarizes the fitted values of these related parameters and compares them with the respective values 
derived from the fully relaxed, spin-resolved DFT+U calculation, which quantitatively agree on the single picometer scale. 

\begin{figure*}[h]
	\includegraphics[width=0.999\textwidth]{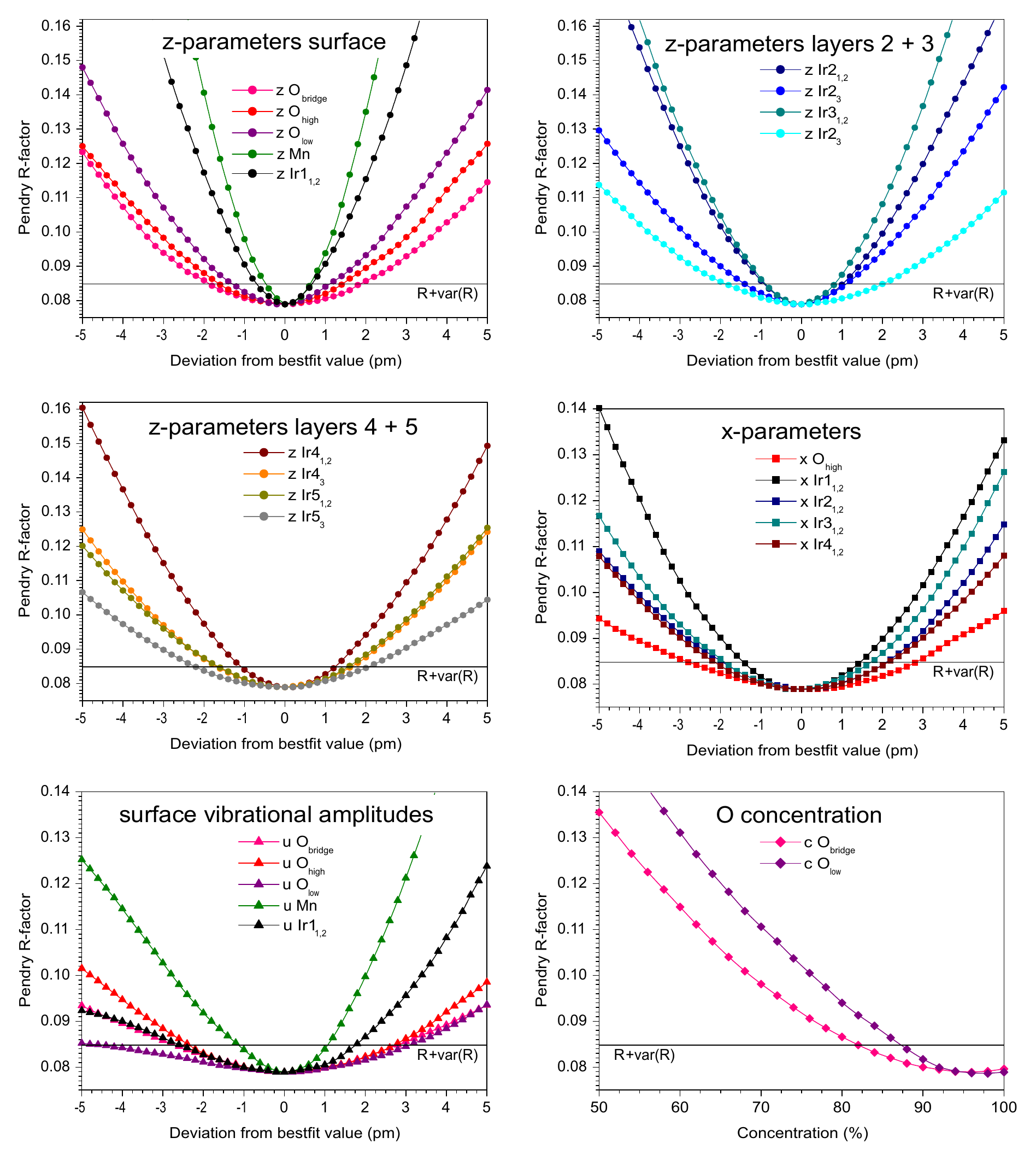}
	\caption{$R$-factor variation of single atom parameters adjusted in the course of the LEED-IV analysis. 
	For each curve the respective parameter was varied only, while all others were held at their best-fit values. 
	The error margins for each parameter are given by the range, where the $R$-factor lies below the variance level $R+var(R) = 0.085$.} \label{fig:errors}
\end{figure*}

Both the large data basis and even more the very low $R$-factor value also led 
to very small error margins in the single picometer range for the determined parameter values. 
According to Pendry \cite{1980-J.Phys.C-Reliability} these error margins can be estimated 
by the range of parameter values, where the $R$-factor lies below the variance level $R+var(R)$. 
Assuming mutual independence of parameters the corresponding ``error curves'', 
i.e., the course of the $R$-factor versus single parameter variations, are displayed in Fig.\,\ref{fig:errors} for all varied parameters. 
From the error margins determined in such a manner the respective errors for the related quantities 
given in Tab.\,\ref{TableMnO3} were calculated via error propagation.

\vfill
\newpage

\begin{table}[p!]
	\renewcommand{\arraystretch}{1.25}
	\renewcommand{\tabcolsep}{1mm}
	\begin{tabular}{l@{\hspace{7mm}}c@{\hspace{5mm}}c@{\hspace{5mm}}c@{\hspace{5mm}}}
		\hline \hline
                       \multicolumn{4}{c}{\textbf{Ir(100)-3$\times$1-MnO$_3$+O$_\textrm{br}$}}\\
		{Parameter}                       &      LEED                                          & DFT   	& $\Delta_{\text{LEED-DFT}}$ \\[.5ex]
		\hline
		~~~d$_\text{O$_{\text{high}}$}$   & 1.387~$^{\text{+0.017}}_{\text{$-$0.016}}$         & 1.357       			& 0.030          \\
		\hline
		~~~d$_\text{O$_{\text{low}}$}$    & $-$0.687~$^{\text{+0.015}}_{\text{$-$0.013}}~~$    & $-$0.689~~   			& $-$0.002~~     \\
		\hline
		~~~d$_\text{O$_{\text{br}}$}$ & 1.439~$^{\text{+0.020}}_{\text{$-$0.019}}$         & 1.430           		& 0.009          \\
		\hline
		~~~a$_\text{O$_{\text{high}}$-O$_{\text{high}}$}$   & 2.583~$^{\text{+0.040}}_{\text{$-$0.040}}$  & 2.598   	& $-$0.015~~     \\
		\hline
		~~~d$_\text{Mn}$                  & 1.304~$^{\text{+0.009}}_{\text{$-$0.009}}$         & 1.295   				& 0.009          \\
		\hline
		~~~p$_\text{1}$                   & 0.118~$^{\text{+0.020}}_{\text{$-$0.020}}$         & 0.142   				& $-$0.024~~     \\
		\hline
		~~~d$_\text{12}$                  & 1.950~$^{\text{+0.013}}_{\text{$-$0.012}}$         & 1.939   				& 0.011          \\
		\hline
		~~~p$_\text{2}$                   & $-$0.222~$^{\text{+0.028}}_{\text{$-$0.030}}~~$    & $-$0.241~~     		& $-$0.019~~     \\
		\hline
		~~~b$_\text{2}$                   & $-$0.015~$^{\text{+0.014}}_{\text{$-$0.020}}~~$    & $-$0.031~~   	 		& $-$0.016~~     \\
		\hline
		~~~d$_\text{23}$                  & 1.919~$^{\text{+0.016}}_{\text{$-$0.016}}$         & 1.908       			& 0.011          \\
		\hline
		~~~p$_\text{3}$                   & $-$0.026~$^{\text{+0.027}}_{\text{$-$0.024}}~~$    & $-$0.031~~   		 	& $-$0.005~~     \\
		\hline
		~~~b$_\text{3}$                   & 0.008~$^{\text{+0.022}}_{\text{$-$0.020}}$         & 0.004      		  	& 0.004          \\
		\hline
		~~~d$_\text{34}$                  & 1.934~$^{\text{+0.019}}_{\text{$-$0.019}}$         & 1.918     		    	& 0.016          \\
		\hline
		~~~p$_\text{4}$                   & 0.004~$^{\text{+0.031}}_{\text{$-$0.031}}$         & $-$0.009~~     	 	& 0.013          \\
		\hline	
		~~~b$_\text{4}$                   & $-$0.018~$^{\text{+0.019}}_{\text{$-$0.021}}~~$    & $-$0.017~~     	 	& 0.001          \\
		\hline
		~~~d$_\text{45}$                  & 1.919~$^{\text{+0.023}}_{\text{$-$0.022}}$         & 1.920        			& $-$0.001~~     \\
		\hline
		~~~b$_\text{5}$                   & 0.002~$^{\text{+0.026}}_{\text{$-$0.027}}$         & 0.000        			& 0.002          \\
		\hline
		~~~d$_\text{5b}$                  & 1.922~$^{\text{+0.017}}_{\text{$-$0.018}}$         & 1.917      		  	& 0.005          \\
		\hline
		~~~u$_\text{O$_{\text{high}}$}$   & 0.112~$^{\text{+0.027}}_{\text{$-$0.024}}$         & ---         		 	& ---            \\
		\hline	
		~~~u$_\text{O$_{\text{low}}$}$    & 0.061~$^{\text{+0.031}}_{\text{$-$0.045}}$         & ---         		   	& ---            \\
		\hline
		~~~u$_\text{O$_{\text{br}}$}$ & 0.104~$^{\text{+0.028}}_{\text{$-$0.027}}$         & ---          		 	& ---            \\
		\hline
		~~~u$_\text{Mn}$                  & 0.095~$^{\text{+0.011}}_{\text{$-$0.011}}$         & ---         		  	& ---            \\
		\hline
		~~~u$_\text{Ir1}$                 & 0.061~$^{\text{+0.026}}_{\text{$-$0.018}}$         & ---         		  	& ---            \\
		\hline
		~~~c$_\text{O$_{\text{low}}$}$    & 1.00~$^{\text{+0.00}}_{\text{$-$0.14}}$            & ---         		  	& ---            \\
		\hline
		~~~c$_\text{O$_{\text{br}}$}$ & 0.96~$^{\text{+0.04}}_{\text{$-$0.14}}$            & ---         		  	& ---            \\
		\hline \hline
	\end{tabular}
	\vspace{5mm}
	
\caption{\label{TableMnO3} Compilation of experimentally determined structural parameters (``LEED'') 
		as defined in Fig.\,\ref{fig:pardef} for the highly oxidized 3$\times$1-MnO$_3$+O$_\textrm{br}$ chain phase on Ir(100) 
		and corresponding values from spin-resolved DFT+U (``DFT'') as well as the their differences $\Delta_{\text{LEED-DFT}}$. 
		DFT values were scaled by a factor 0.9883 to account for the slightly different DFT lattice parameter. 
		All values are given in {\AA}ngstr\"{o}m. Error margins for the ``LEED'' parameters were derived 
		from single atom uncertainty ranges displayed in Fig.\,\ref{fig:errors} via error propagation. 
		Also given are the LEED best-fit values for vibrational amplitudes of outermost atoms and the fitted concentration 
		of O$_{\text{low}}$ and O$_{\text{br}}$ atoms (c$_\text{O$_{\text{low}}$}$ and ~c$_\text{O$_{\text{br}}$}$).} \centering
\end{table}

\newpage

\newpage

\section*{III. $R$-factor comparison for 1/6 ML M\lowercase{n} preparation}

\begin{figure*}[p!]
	\includegraphics[width=0.75\textwidth]{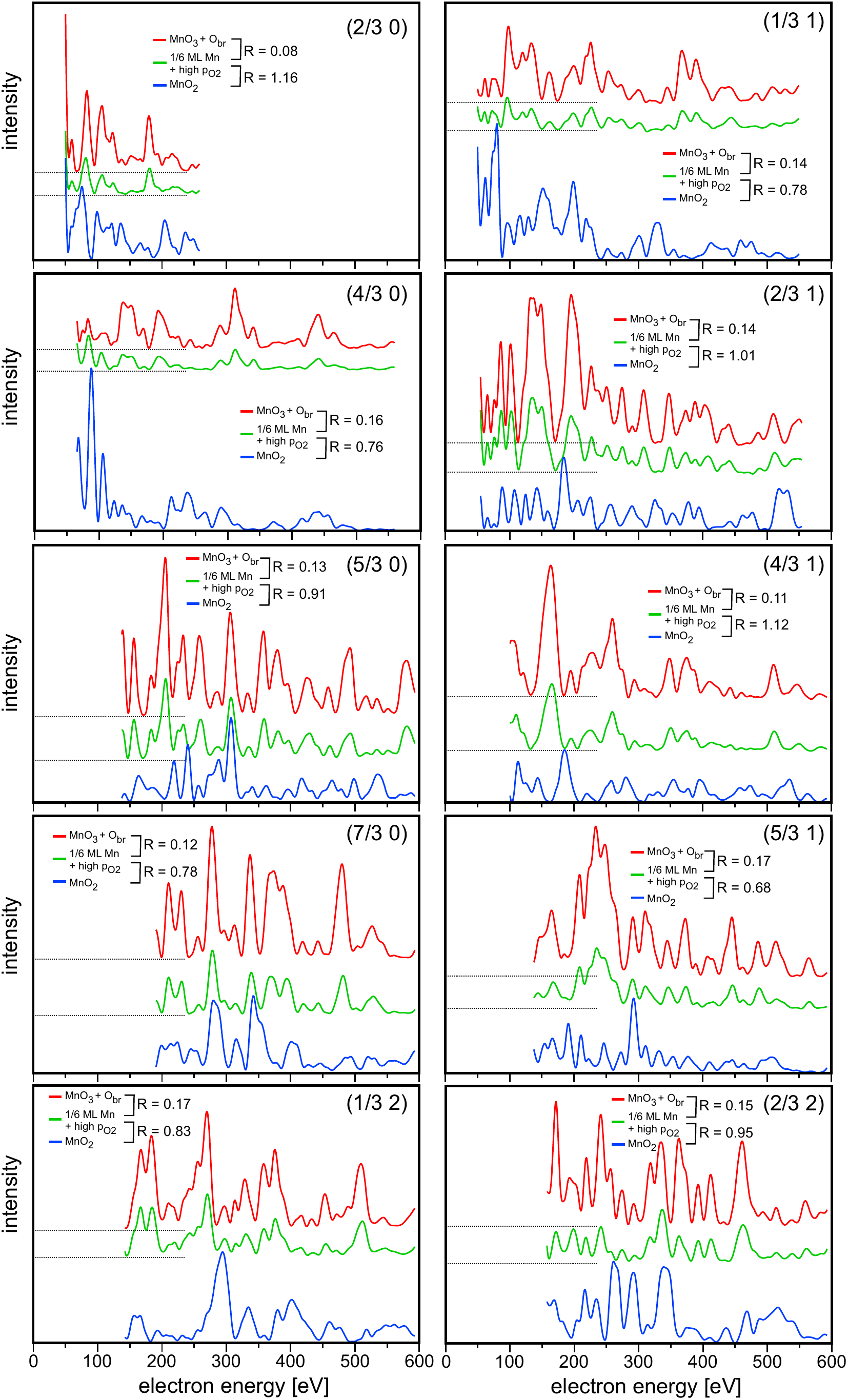}
\caption{$R$-factor analysis of experimental spectra for different surface preparations. 
		Each panel is labeled by the (h k) index of the respective scattering vector. 
		Curves are offset for clarity. 
		Top curves correspond to the \MnOO+O$_\textrm{br}$ phase prepared in NO$_2$, 
		middle curves to the preparation with 1/6 ML Mn oxidized in $10^{-6}$\,mbar O$_2$, 
		lower curves to the fully developed \MnO\ phase.}
	\label{fig:FigS2}
\end{figure*}

The $R$-factor also allows to express the similarity between experimental spectra. 
Here we compare the experimental spectra of the highly oxidized preparation with 1/6 ML Mn on Ir(100) 
with those of the Ir(100) fully covered by the \MnOO+O$_\textrm{br}$ phase or by the \MnO\ phase.
Since with 1/6 ML Mn only half the surface is covered with MnO$_x$ chains and the other is covered with Ir(100)-$(2\times 1)$-O 
we only compare those beams that belong to the \threebyone\ surface cell and not to the $(2\times 1)$. 
This also causes the absolute intensity of these beams to be only half of the corresponding fully covered surface. 
As becomes apparent from the collection in Fig.\,\ref{fig:FigS2}, the highly oxidized 1/6 ML preparation 
corresponds closely to that of the fully covered \MnOO+O$_\textrm{br}$ surface structure 
with an overall $R$-factor of $R=0.123$ using a data basis of $\Delta E = 4480\,\textrm{eV}$. 
In contrast, the comparison with the \MnO\ chain phase yields an $R$-factor as high as $R=0.806$. 
Note that an $R$-factor of $R = 1.0$ characterizes completely uncorrelated intensity spectra. 
We thus can safely rule out the existence of a significant share 
of low-oxidized \MnO\ chains at the surface under the chosen preparation conditions.

\newpage

\section*{IV. Phase stability analysis by DFT}

\begin{figure*}[h]
	\includegraphics[width=0.9\textwidth]{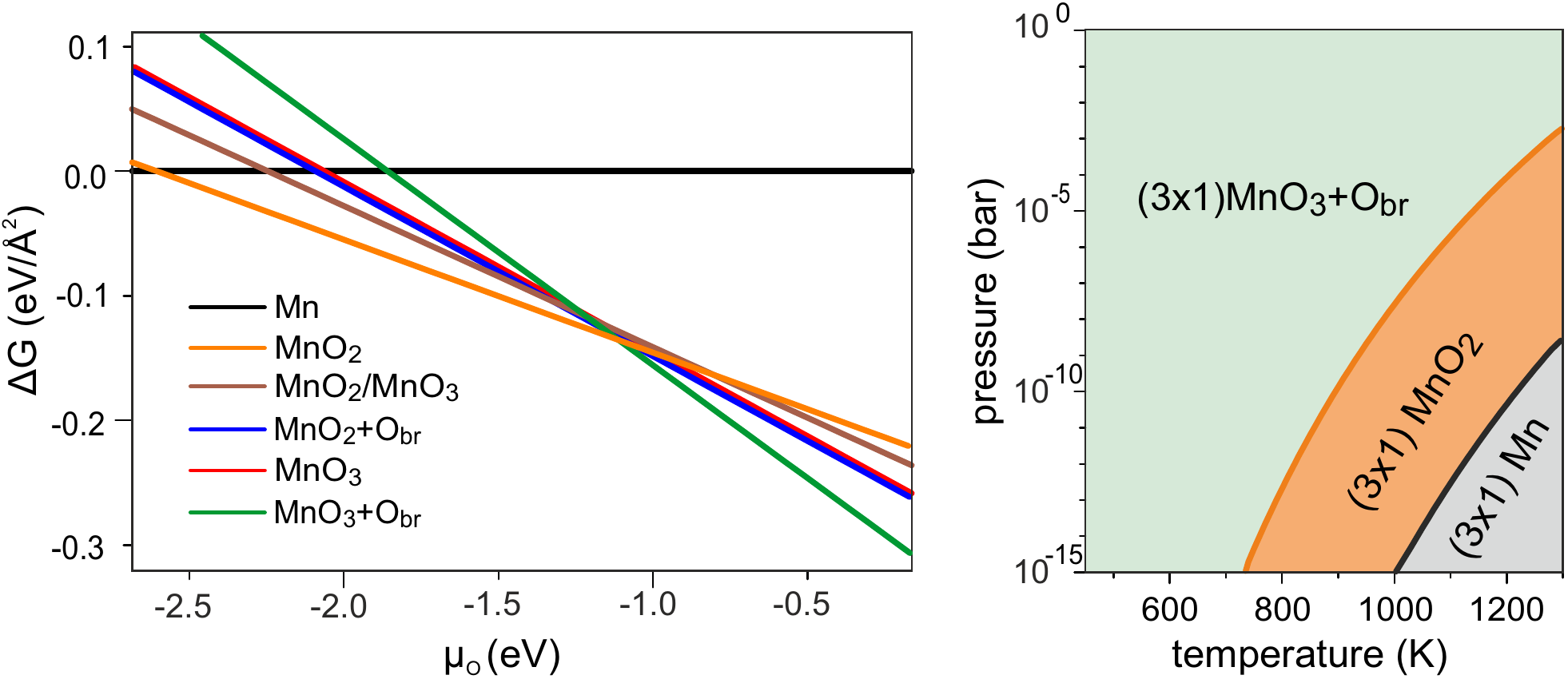}
	\caption{Stability analysis (left) and phase diagram (right) as obtained from spin-polarized DFT+U calculations. 
	Explanation see text.}     \label{fig:PhaseDiagram}
\end{figure*}

From our DFT calculations using a $(6\times 2)$ surface unit cell we obtained the binding energy per oxygen atom $E_B(\sigma)$ 
for a particular chain structure $\sigma$ with respect to a purely metallic \threebyone Mn structure \cite{2017-PRB-Monatomic} from:
\begin{equation}
	E_B (\sigma) =  - \frac{1}{N_{\text{O}}} \left[ E_\sigma - E_\text{\threebyone Mn}-\frac{N_{\text{O}}}{2} E_{\text{O}_2} \right]\ \ ,
\end{equation}
where $E_\sigma$ and $E_\text{\threebyone Mn}$ are the total energies of the structure $\sigma$ 
and the Ir(100)-\threebyone Mn reference structure respectively, $N_{\text{O}}$ is the number of oxygen atoms in the unit cell, 
and $E_{\text{O}_2}$ is the total energy of triplet O$_2$. 
For a given oxygen chemical potential $\mu_{\text{O}}$ the change in Gibbs surface free energy $\Delta G$ 
with respect to the oxygen free Ir(100)-\threebyone Mn may be written as \cite{Rogal2007}:
\begin{equation}
	\Delta G =  - \frac{N_{\text{O}}}{A} [E_{\text{B}}(\sigma) +
	\mu_{\text{O}}] \ \label{free_energy}\ \ ,
\end{equation} 
where $A$ is the surface area of the $(6\times 2)$ cell. 
Negative $\Delta G$ designates a phase $\sigma$ that is thermodynamically more stable 
at a particular $\mu_{\text{O}}$ than the \threebyone Mn phase. 
We note that the reference phase may in principle be prepared by reduction in hydrogen \cite{2017-PRB-Monatomic}. 
Equivalently to a plot $\Delta G$ as function of $\mu_{\text{O}}$ the dependence of the latter on pressure $p$ and temperature $T$,
\begin{equation}
	\mu_{\text{O}} = \mu_{\text{O}}^0 + \frac{1}{2} k_\text{B}T
	\ ln\left( \frac{\text{p}}{\text{p}^0}\right) \ \ ,
	\label{chempot}
\end{equation}
may be used to plot a $p$-$T$ phase diagram where the boundaries designate the points ($p$,$T$) 
when the corresponding state is the most favorable one, i.e., $\mu_{\text{O}} (p,T) = - E_B(\sigma)$. 
We interpolated the reference state $\mu_{\text{O}}^0$ from \cite{Reuter2001} and used $p^0 = 1\,\textrm{bar}$ as reference pressure. 
The structures $\sigma$ we analyzed were (in order of increasing oxygen content): 
the \threebyone \MnO, the phase with alternating \MnO\ and \MnOO\ chains, the \threebyone \MnO+O$_\textrm{br}$, 
the \MnOO, and finally the \MnOO+O$_\textrm{br}$.   The result is shown in Fig.\,\ref{fig:PhaseDiagram}. 
The \threebyone Mn is only stable for very low $\mu_{\text{O}}$, the \MnO\ phase is the most favorable one 
at intermediate $\mu_{\text{O}}$ due to its largest $E_B$, whereas---due to its largest oxygen content---the 
\MnOO+O$_\textrm{br}$ is the most stable phase at high $\mu_{\text{O}}$. 
The mixed \MnO /\MnOO\ phase is in principle more stable than the \MnOO\ or \MnO+O$_\textrm{br}$ phases, 
but both are thermodynamically less stable than the pure \MnO\ phase. 
The resulting phase diagram in the right panel of Fig.\,\ref{fig:PhaseDiagram} demonstrates 
that the \MnOO+O$_\textrm{br}$ converts to the \MnO\ phase in UHV above approximately 800\,K, 
quite in agreement with experimental observations.

\newpage

\bibliography{Bibliography_v23}